\documentclass[prd,showpacs,amsmath,amssymb,twocolumn]{revtex4}
\usepackage{graphicx,color}
\begin{document}

\title{Study on $B_s\to D_{sJ}(2317,2460)l\bar{\nu}$ Semileptonic Decays
in the CQM Model} \vspace{3mm}

\author{Shu-Min Zhao$^1$}\author{Xiang Liu$^2$}\email{xiangliu@pku.edu.cn}
\author{Shuang-Jiu Li$^1$} \affiliation{1. Department of
Physics and Technology, Hebei University, Baoding 071002, China}

\affiliation{2. Department of Physics, Peking University, Beijing
100871, China}

\vspace*{1.0cm}

\date{\today}

\begin{abstract}
Assuming $D^*_{sJ}(2317)$ and $D_{sJ}(2460)$ to be the $(0^+,1^+)$
chiral partners of regular $D_{s}(1968)$ and $D^{*}_{s}(2112)$, we
calculate the semileptonic decays of $B_s$ to $D_s(1968)$,
$D^*_s(2112)$, $D_{sJ}^{*}(2317)$, $D_{sJ}(2460)$ in terms of the
Constituent Quark Meson (CQM) model. The large branching ratios of
the semileptonic decays of $B_s$ to $D_{sJ}^{*}(2317)$ and
$D_{sJ}(2460)$ indicate that those two semileptonic decays should be
seen in future experiments.
\end{abstract}


\maketitle
%\newpage

\section{Introduction}

The discoveries of exotic mesons $D^*_{sJ}(2317)$ and $D_{sJ}(2460)$
\cite{2317,Belle,CLEO} whose spin-parity structure are respectively
$0^+$ and $1^+$, have attracted great interests of both theorists
and experimentalists of high energy physics. $D_{sJ}^*(2317)$ and
$D_{sJ}(2460)$ are supposed to be $(0^+,1^+)$ chiral partners of
$D_s$ and $D_s^*$ \cite{Bardeen} i.e. p-wave excited states of $D_s$
and $D_s^*$ \cite{Chao}. Beveren and Rupp suggested that
$D_{sJ}^*(2317)$ and $D_{sJ}(2460)$ are made up of $c$ and $\bar{s}$
by studying the mass spectra \cite{Beverenn1}. With the QCD spectral
sum rules, Narison calculated the masses of $D_{sJ}^*(2317)$ and
$D_{sJ}(2460)$ by assuming them as quark-antiquark states and
obtained results which are consistent with the experiment data
within a wide error range \cite{Narison}. Very recently, considering
the contribution of DK continuum in QCD sum rules, Dai et al.
obtained the mass of $D_{sJ}^{*}(2317)$ which is consistent with the
mass given by the experiments \cite{Dai}. Meanwhile, some authors
suggested that $D_{sJ}^*(2317)$ and $D_{sJ}(2460)$ may be of
four-quark structure \cite{Chen,Cheng,T. Barnes}. Thus one needs to
use various theoretical approaches to clarify the mist of the
structures of $D_{sJ}^*(2317)$ and $D_{sJ}(2460)$. The studies of
the productions and decays of $D_{sJ}^*(2317)$ and $D_{sJ}(2460)$
are very interesting topics.

The semileptonic decays of $B_{s}$ are one of the ideal platforms to
study the productions of $D_{sJ}^*(2317)$ and $D_{sJ}(2460)$.
Especially the Large Hadron Collider (LHC) will be run in the coming
2007, which can produce large amounts of data of $B_s$. Thus the
measurements of $B_{s}\to D_{sJ}(2317,2460)l\bar{\nu}$ would be
realistic. In Ref. \cite{huang}, author calculated $B_{s}\to
D_{sJ}(2317,2460)l\bar{\nu}$ decays by the QCD sum rules in HQET.
Recently, authors of Ref. \cite{aliev1,aliev2} completed the
calculations of $B_{s}\to D_{sJ}(2317,2460)l\bar{\nu}$ semileptonic
decays in the QCD sum rules and obtained large branching ratios.
However, the results obtained by Ref. \cite{aliev1,aliev2} are one
order smaller than those given by Ref. \cite{huang}. Thus studies on
$B_{s}\to D_{sJ}(2317,2460)l\bar{\nu}$ with other plausible models
would be helpful. It not only deepen our understanding about the
properties of these states but also test the reliability of models
which are applied to calculate the semileptonic decays.

In this work, we study the $B_{s}\to D_{s}(1968)l\bar{\nu}$,
$B_{s}\to D_{s}^{*}(2112)l\bar{\nu}$ and $B_{s}\to
D_{sJ}(2317,2460)l\bar{\nu}$ semileptonic dedays in terms of the
Constituent Quark Meson (CQM) model. In order to complete the
calculations of the semileptonic decays of $B_s$, we firstly base on
the assumption that ($D_s(1968),D^*_s(2112)$) with spin-parity
($0^-,1^-$) and ($D^*_{sJ}(2317),D_{sJ}(2460)$) with spin-parity
($0^+,1^+$) can be respectively categorized as $H(0^-,1^-)$ and
$S(0^+,1^+)$ doublets in HQET.

CQM model was proposed by Polosa et al. \cite {CQM} and has been
well developed later based on the work of Ebert et al.
\cite{feldmann} (See the Ref. \cite{CQM} for a review). The model is
based on an effective Lagrangian which incorporates the flavor-spin
symmetry for heavy quarks with the chiral symmetry for light quarks.
Employing the CQM model to study the phenomenology of heavy meson
physics, reasonable results have been achieved
\cite{application,parameter}. Therefore, we believe that the model
is applicable to our processes and expect to get relatively reliable
conclusion.

This paper is organized as follow: After the introduction, in Sect.
II, we formulate the semileptonic decays of $B_s$ to $D_s(1968)$,
$D_s^*(2112)$, $D_{sJ}^*(2317)$ and $D_{sJ}(2460)$. The numerical
results along with all the input parameters are presented in Sect
III. Section IV is devoted to the discussion and the conclusion.
Some detailed expressions are collected in the appendix.

\section{formulation}

For the convenience of readers, we give a brief introduction of the
CQM model \cite{CQM}. The model is relativistic and based on an
effective Lagrangian which combines the HQET and the chiral symmetry
for light quarks
\begin{eqnarray}
\mathcal{L}_{CQM}&=&\bar{\chi}[\gamma\cdot(i\partial+V)]\chi+\bar{\chi}\gamma\cdot
A\gamma_5\chi-m_q\bar{\chi}\chi
\nonumber\\&&+\frac{f^2_{\pi}}{8}Tr[\partial^{\mu}\Sigma\partial_{\mu}\Sigma^+]+
\bar{h}_{v}(iv\cdot\partial)h_v\nonumber\\
&&-[\bar{\chi}(\bar{H}+\bar{S}+i\bar{T}^{\mu}\frac{\partial_{\mu}}{\Lambda})h_{v}+h.c.]
\nonumber\\&&+\frac{1}{2G_3}
Tr[(\bar{H}+\bar{S})(H-S)]+\frac{1}{2G_4}Tr[\bar{T}^{\mu}T_{\mu}].\nonumber
\end{eqnarray}
where the fifth term is the kinetic term of heavy quarks with
$\slash\!\!\!vh_{v}=h_v$; $H$ and $S$ denote the super-fields
corresponding to doublets ($0^-,1^-$) and ($0^+,1^+$) respectively.
The explicit matrix representations of $H$ and $S$ read as
\cite{luke}
\begin{eqnarray}
H&=&\frac{1+ v\!\!\!\slash}{2}[P_{\mu}^*\gamma^{\mu}-P\gamma_5],\\
S&=&\frac{1+v\!\!\!\slash}{2}[P_{1\mu}^{*'}\gamma^{\mu}\gamma_5-P_{0}],
\end{eqnarray}
where $P$, $P^{*\mu}$, $P_{0}$ and $P_{1}$ are the annihilation
operators of pseudoscalar, vector, scalar and axial vector mesons
which are normalized as {\small
\begin{eqnarray*}
\langle0|P|M(0^-)\rangle&=&\sqrt{M_{H}},\;\;
\langle0|P^{*\mu}|M(1^-)\rangle=\sqrt{M_{H}}\epsilon^{\mu},\\
\langle0|P_{0}|M(0^+)\rangle&=&\sqrt{M_{S}}\gamma_{5},\;
\langle0|P_{1}^{*\mu}|M(1^+)\rangle=\sqrt{M_{S}}\gamma_{5}\epsilon^{\mu}.
\end{eqnarray*}}
$T$ is the super-field corresponding to the doublet ($1^+,2^+$)
{\small
\begin{eqnarray}
T^{\mu}=\frac{1+
v\!\!\!\slash}{2\sqrt{2}}\bigg[P_2^{*\mu\nu}\gamma_{\nu}-\sqrt{\frac{3}{2}}P_{1\nu}^*\gamma_5\bigg(g^{\mu\nu}-
\frac{1}{3}\gamma^{\nu}(\gamma^{\mu}-v^{\mu})\bigg)\bigg].
\end{eqnarray}}
$\chi=\xi q (q=u,d,s)$ is the light quark field and
$\xi=e^{\frac{iM}{f_{\pi}}}$, and $M$ is the octet pseudoscalar
matrix. We also have
\begin{eqnarray}
&&V^{\mu}=\frac{1}{2}(\xi^{\dag}\partial^{\mu}\xi+\xi\partial^{\mu}\xi^{\dag}),\\
&&A^{\mu}=\frac{-i}{2}(\xi^{\dag}\partial^{\mu}\xi+\xi\partial^{\mu}\xi^{\dag}).
\end{eqnarray}
Because the spin-parity of $D_s(1968)$ and $D_s^*(2112)$ are
respectively $0^-$ and $1^-$, $D_s(1968)$ and $D_s^*(2112)$ can be
embedded into the $H$-type doublet ($0^-,1^-$), whereas
$D_{sJ}^*(2317)$ and $D_{sJ}(2460)$ belong to the $S$-type doublet
($0^+,1^-$). Thus we can calculate the semileptonic decays of $B_s$
to $D_s(1968)$, $D_s^*(2112)$, $D_{sJ}^*(2317)$ and $D_{sJ}(2460)$.

\subsection{The calculations of $B_s\rightarrow D_{s}(1968)l\bar{\nu}$
and $B_s\rightarrow D^*_{s}(2112)l\bar{\nu}$ in the CQM model.}

The four fermion operator of $b\rightarrow c+l\bar{\nu}$ which is
relevant to the semileptonic decays of $B_s$ to $D_s$ mesons reads
as \cite{Xiang nan}
\begin{eqnarray}
\mathcal{O}=\frac{G_FV_{cb}}{\sqrt{2}}\bar{c}\gamma^{\mu}(1-\gamma_5)b\bar{\nu}\gamma_{\mu}(1-\gamma_5)l.
\end{eqnarray}

The transition amplitudes of $B_s\rightarrow D_{s}(1968)l\bar{\nu}$
and $B_s\rightarrow D^*_{s}(2112)l\bar{\nu}$ can be written as
\begin{eqnarray}
&&\mathcal{M}=\langle
D_{s}^{(*)}l\bar{\nu}|\mathcal{O}|B_{s}\rangle\nonumber\\&&=\frac{G_FV_{cb}}{\sqrt{2}}\langle
D_{s}^{(*)}|\bar{c}\gamma^{\mu}(1-\gamma_5)b|B_{s}\rangle \langle
l\bar{\nu}|\bar{\nu}\gamma_{\mu}(1-\gamma_5)l|0\rangle,\nonumber\\\label{la}
\end{eqnarray}
where the hadronic matrix element is related to non-perturbative QCD
effects. In the HQET symmetries, the hadronic matrix element can be
expressed with the following form
\begin{eqnarray}
&&\langle D_{s}(v')|\bar{c}\gamma_{\mu}(1-\gamma_5)b|B_{s}(v)\rangle
\nonumber\\&&=\sqrt{M_{B_{s}}M_{D_{s}}}(v_{\mu}+v'_{\mu})\label{factor1}\xi(\omega),\\
&&\langle
D_{s}^*(v',\epsilon)|\bar{c}\gamma_{\mu}(1-\gamma_5)b|B_{s}(v)\rangle
\nonumber\\&&=\sqrt{M_{B_{s}} M_{D_{s}^*}}[i\varepsilon
_{\mu\nu\alpha\beta}\epsilon^{*\nu}v'^{\alpha}v^{\beta}
-(1+\omega)\epsilon_{\mu}^{*}\nonumber\\&&+(\epsilon_{\mu}^{*}\cdot
v)v'_{\mu}]\xi(\omega)\label{factorn1},
\end{eqnarray}
where $\omega=v\cdot v'$. In HQET, $\xi(\omega)$ is the Isgur-Wise
function which is a dimensionless probability function. In the
following, the central task is how to extract the Isgur-Wise
function in the calculation of the CQM model.

In the CQM model, the Feynman diagram corresponding to the hadronic
matrix element $\langle
D_{s}^{(*)}|\bar{c}\gamma^{\mu}(1-\gamma_5)b|B_{s}\rangle$ can be
depicted in Fig. 1.
\begin{figure}[htb]
\begin{center}
\scalebox{0.8}{\includegraphics{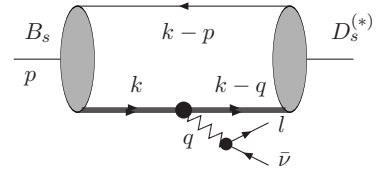}}
\end{center}
\caption{The Feynman diagrams depict the decays of $B_{s}\rightarrow
D^{(*)}_{s}l\bar{\nu}$. The thick-line denotes the heavy quark
propagator.} \label{diagram}
\end{figure}

According to the CQM model \cite{CQM}, the couplings of
$D_{s}(1968)$, $D_{s}^{*}(2112)$ and $B_{s}$ with light and heavy
quarks are expressed as
\begin{eqnarray}
&&\frac{1+ v\!\!\!\slash}{2}\sqrt{Z_{H}M_{D_{s}}}\;\gamma_{5},\\
&&\frac{1+ v\!\!\!\slash}{2}\sqrt{Z_{H}M_{D^{*}_{s}}}\;\epsilon\!\!\!\slash,\\
&&\frac{1+ v\!\!\!\slash}{2}\sqrt{Z_{H}M_{B_{s}}}\;\gamma_{5},
\end{eqnarray}
where $\epsilon$ denotes the polarization vector of $D^{*}(2112)$.
The concrete expression of $Z_{H}$ is given in \cite{CQM} as
\begin{eqnarray}
Z_{H}^{-1}&=&(\Delta_{H}+m_{s})\frac{\partial
\mathcal{I}_{3}(\Delta_{H})}{\partial\Delta_{H}}+ \mathcal{I}_{3}(\Delta_{H}),\label{ZH}\\
\mathcal{I}_{3}(a)&=&\frac{iN_c}{16\pi^4}\int^{1/\mu^2}_{1/\Lambda^2}\frac{dy}{y^{3/2}}
\exp[-y(m_{s}^{2}-a^2)]\nonumber\\&&\times(1+\mathrm{erf}(a\sqrt{y})),
\end{eqnarray}
where erf is the error function. $m_s$ is the mass of the $s$ quark.

Now we can write out the hadronic transition matrix element
\begin{eqnarray}
&&\langle D_{s}^{(*)}|\bar{c}\gamma^{\mu}(1-\gamma_5)b|B_{s}\rangle
\nonumber\\&=&\frac{(i^5)}{4}\sqrt{M_{B_{S}}M_{D_{s}^{(*)}}}\;Z_{H}
N_{c}\nonumber\\&&\times\int^{reg}\frac{\mathrm{d}^{4}k}{(2\pi)^{4}}
\frac{\mathrm{Tr}\big[(k\!\!\!\slash+m)\;\Gamma\;(1+
v\!\!\!\slash')\gamma^{\mu}(1+
v\!\!\!\slash)\gamma^5\big]}{(k^2-m_{s}^2)(v'\cdot
k+\Delta_{H})(v\cdot k+\Delta_{H})}\nonumber\\\label{haha}
\end{eqnarray}
with $N_{c}=3$, where $\Gamma$ should be taken as $\gamma_{5}$ and
$\epsilon\!\!\!\slash$ corresponding to $D_{s}$ and $D_{s}^{*}$
respectively.

One omits the technical details in the text for saving space and
finally gets the Isgur-Wise function $\xi(\omega)$ by comparing the
results of eq. (\ref{haha}) to eqs. (\ref{factor1}) and
(\ref{factorn1})
\begin{eqnarray}
\xi(\omega)&=&Z_H\bigg[\frac{2}{1+\omega}\mathcal{I}_3(\Delta_H)+\bigg(m_{s}+\frac{2\Delta_H}{1+\omega}\bigg)
\nonumber\\&&\times \mathcal{I}_5(\Delta_H,\Delta_H,\omega)\bigg],
\end{eqnarray}
where the definitions of $\mathcal{I}_{1,5}$ are listed in the
appendix.

By eqs. (\ref{la})-(\ref{factorn1}), we obtain the explicit
expressions of the semileptonic decays $B_s\rightarrow
D_{s}(1968)l\bar{\nu}$ and $B_s\rightarrow D^*_{s}(2112)l\bar{\nu}$.
\begin{eqnarray}
&&\mathrm{d}\Gamma(B_s\rightarrow D_{s}(1968)l\bar{\nu}
)\nonumber\\&=&\frac{G_F^2|V_{cb}|^2}{48\pi^3}(M_{B_{s}}+M_{D_{s}})^2M_{D_{s}}^3(\omega^2-1)^{3/2}\xi^2(\omega)d\omega\label{k1968},\nonumber\\
&&\mathrm{d}\Gamma(B_s\rightarrow
D^*_{s}(2112)l\bar{\nu})\nonumber\\&=&\frac{G_F^2|V_{cb}|^2}{48\pi^3}(M_{B_{s}}-M_{D_{s}}^*)^2M_{D_{s}}^{*3}
\sqrt{\omega^2-1}(\omega+1)^{2}\nonumber\\
&&\bigg[1+\frac{4\omega}{(1+\omega)}\frac{q^2}{(M_{B_s}-M_{D_s}^*)^2}\bigg]\xi(\omega)^2d\omega\label{k2112},
\end{eqnarray}
where $q^2=M_B^2+M_D^2-2M_BM_D\omega$.

\subsection{The calcualtions of $B_s\rightarrow D^*_{sJ}(2317)l\bar{\nu}$ and $B_s\rightarrow D_{sJ}(2460)l\bar{\nu}$
in the CQM model}

The hadronic matrix elements of the decays $B_s\rightarrow
D^*_{sJ}(2317)l\bar{\nu}$ and $B_s\rightarrow
D_{sJ}(2460)l\bar{\nu}$ in HQET read as \cite{isgur}
\begin{eqnarray}
&&\langle D_{sJ}^{*}(2317)|\bar{c}\gamma_{\mu}(1-\gamma_5)b|B_{s}(v)\rangle\nonumber\\
&=&2\sqrt{M_{B_{s}}M_{D_{sJ}}(2317)}(v'_{\mu}-v_{\mu})\zeta(\omega)\label{factor2},\\
&&\langle
D_{sJ}(2460)(v',\epsilon')|\bar{c}\gamma_{\mu}(1-\gamma_5)b|B_{s}(v)\rangle\nonumber\\&=&
\sqrt{M_{B_{s}}M_{D_{sJ}(2460)}}
\bigg\{2i\varepsilon_{\mu\alpha\beta\gamma}\epsilon^{*\alpha}v^{'\beta}v^{\gamma}\nonumber\\
&&+2[(1-\omega)\epsilon^*_{\mu}+(\epsilon^*\cdot
v)v_{\mu}']\bigg\}\zeta(\omega)\label{factor3},
\end{eqnarray}
where $\zeta(\omega)$ denote the the Isgur-Wise function.

We use the same treatment in subsection A to get $\zeta(\omega)$. In
the CQM model, the couplings of $D_{sJ}^{*}(2317)(0^+)$ and
$D_{sJ}(2460)(1^{+})$ with light and heavy quarks are respctively
\begin{eqnarray}
&&\frac{1+ v\!\!\!\slash}{2}\sqrt{Z_{S}M_{D_{sJ}^{*}(2317)}},\\
&&\frac{1+
v\!\!\!\slash}{2}\sqrt{Z_{S}M_{D_{sJ}(2460)}}\;\epsilon\!\!\!\slash'\gamma_{5}
\end{eqnarray}
with
\begin{eqnarray}
Z_{S}^{-1}&=&(\Delta_{S}+m_{s})\frac{\partial
\mathcal{I}_{3}(\Delta_{S})}{\partial\Delta_{S}}+
\mathcal{I}_{3}(\Delta_{S}),\label{ZS}
\end{eqnarray}
where $\epsilon'$ is the polarization vector of $D_{sJ}(2460)$.

Thus, in the CQM modle, the hadronic transition matrix elements of
$\langle
D_{sJ}^{*}(2317)|\bar{c}\gamma_{\mu}(1-\gamma_5)b|B_{s}(v)\rangle$
and $\langle
D_{sJ}(2460)(v',\epsilon')|\bar{c}\gamma_{\mu}(1-\gamma_5)b|B_{s}(v)\rangle$
can be obtained by replacing $\Gamma$ in eq. (\ref{haha}) with $1$
and $\epsilon\!\!\!\slash'\gamma_{5}$ respectively. Meanwhile,
$Z_{H}$ should be replaced by $\sqrt{Z_{H}Z_{S}}$.

Finally we get the Isgur-Wise function $\zeta(\omega)$ with the
following form
\begin{eqnarray}
\zeta(\omega)&=&\frac{\sqrt{Z_H
Z_S}}{2(1-\omega)}[\mathcal{I}_3(\Delta_S)-\mathcal{I}_3(\Delta_H)\nonumber\\&&+(\Delta_H-\Delta_S+m_{s}(1-\omega))
\mathcal{I}_5(\Delta_H,\Delta_S)].\nonumber\\
\end{eqnarray}

Using eqs. (\ref{la}), (\ref{factor2}) and (\ref{factor3}), we
deduce the decay widths of $B_s\rightarrow D^*_{sJ}(2317)l\bar{\nu}$
and $B_s\rightarrow D_{sJ}(2460)l\bar{\nu}$
\begin{eqnarray}
&&\mathrm{d}\Gamma(B_s\rightarrow
D^*_{sJ}(2317)l\bar{\nu})\nonumber\\&
=&\frac{G_F^2|V_{cb}|^2}{12\pi^3}
M_{D_{sJ}^{*}(2317)}^3(M_{B_{s}}-M_{D_{sJ}^{*}(2317)})^2\nonumber\\
&&\times(\omega^2-1)^{3/2}\zeta^2(\omega)d\omega \label{k2317},\\
&&\mathrm{d}\Gamma(B_s\rightarrow
D_{sJ}(2460)l\bar{\nu})\nonumber\\&&=\frac{G_F^2|V_{cb}|^2}{12\pi^3}M_{D_{sJ}(2460)}^3\sqrt{(\omega^2-1)}
 \zeta^2(\omega)\nonumber\\
 &&\times\bigg[(M_{B_{s}}^2+M_{D_{sJ}(2460)}^2)(5\omega^2-6\omega+1)\nonumber\\&&
 -2M_{B_{s}}M_{D_{sJ}(2460)}(4\omega^3-5\omega^2+2\omega-1)\bigg]
 d\omega\label{k2460}.\nonumber\\
\end{eqnarray}

\section{Numerical results}

The semileptonic decays $B^{+}\to \bar{D}^{(*)0}l^{+}\nu$ and
$B^{0}\to D^{(*)-}l^{+}\nu$ are measured well \cite{PDG}. The
results calculated by the CQM model \cite{CQM} and measured by the
experiments are collected in Table \ref{B-Dlnu}. By comparing the
theoretical results with that measured by the experiments, one
believes that the CQM model is applicable to our processes and
expects to get relatively reliable result.
\begin{center}
\begin{table}[htb]
\begin{tabular}{c|cc}\hline
&CQM model\cite{CQM}& Experiment \cite{PDG} \\\hline \hline
$BR[B^{+}\to \bar{D}^{0}l^{+}\nu]$& &$(2.15\pm0.22)\%$\\\cline{1-1}
\cline{3-3}$BR[B^{0}\to
D^{-}l^+\nu]$&\raisebox{2ex}[0pt]{(2.2$\sim$3.0)\%}&$(2.12\pm0.20)\%$\\\hline
$BR[B^{+}\to \bar{D}^{*0}l^{+}\nu]$&
&$(6.5\pm0.5)\%$\\\cline{1-1}\cline{3-3} $BR[B^{0}\to
D^{*-}l^+\nu]$&\raisebox{2ex}[0pt]{(5.9$\sim$7.6)\%}&$(5.35\pm0.20)\%$\\\hline
\end{tabular}\caption{The numerical results are taken from Ref. \cite{CQM}.
\label{B-Dlnu}}
\end{table}
\end{center}

With the formulation derived from the last section, one numerically
evaluate the corresponding decay rates. The input parameters
include: $G_F=1.1664\times 10^{-5}$ GeV$^{-2}$, $V_{cb}=0.043$,
$M_{B_s}=5.3696$ GeV, $M_{D_s}=1.968$ GeV, $M_{D^*_s}=2.112$ GeV,
$M_{D_{sJ}^{*}(2317)}=2.317$ GeV, $M_{D_{sJ}(2460)}=2.46$ GeV
\cite{PDG}. $m_s=0.5$ GeV, $\Lambda=1.25$ GeV, the infrared cutoff
$\mu=0.593$ GeV and $\Delta_{S}-\Delta_{H}=335\pm 35$ MeV
\cite{parameter}.

We present the branching ratios of $B_{s}\to D_{s}(1968)l\bar{\nu}$,
$B_{s}\to D_{s}^{*}(2112)l\bar{\nu}$, $B_{s}\to
D_{sJ}^{*}(2317)l\bar{\nu}$ and $B_{s}\to D_{sJ}(2460)l\bar{\nu}$ in
Table \ref{tab}.
\begin{widetext}
\begin{center}
\begin{table}[h]
%\begin{center}
\renewcommand{\arraystretch}{1}
\tabcolsep=0.1cm
\begin{tabular}{cccccccc}
  \hline\
  $\Delta_H$ (GeV) & $\Delta_S$ (GeV)&  $Z_H$ (GeV)$^{-1}$ &$Z_S$ (GeV)$^{-1}$ &$BR_1$ &$BR_2$ &$BR_3$ &$BR_4$\\
  \hline\hline
  0.5 & 0.86 & 3.99 & 2.02&$2.95\%$&$7.66\%$&$5.71\times10^{-3}$&$8.69\times10^{-3}$ \\
  \hline
  0.6 & 0.91  & 2.69 & 1.47 &$2.86\%$&$7.53\%$&$5.25\times10^{-3}$&$7.91\times10^{-3}$\\
 \hline
 0.7 & 0.97 &1.74 & 0.98&$2.73\%$&$7.49\%$&$4.90\times10^{-3}$&$7.52\times10^{-3}$ \\
 \hline
\end{tabular}\caption{The values of $\Delta_{S}$ and $\Delta_{H}$ are taken from
\cite{parameter}. According to eqs. (\ref{ZH}) and (\ref{ZS}), one
gets the values of $Z_{S}$ and $Z_{H}$. $BR_{1}$, $BR_{2}$, $BR_{3}$
and $BR_{4}$ respectively denote the the branching ratios of
$B_{s}\to D_{s}(1968)l\bar{\nu}$, $B_{s}\to
D_{s}^{*}(2112)l\bar{\nu}$, $B_{s}\to D_{sJ}^{*}(2317)l\bar{\nu}$
and $B_{s}\to D_{sJ}(2460)l\bar{\nu}$.\label{tab}}
%\end{center}
\end{table}
\end{center}
\end{widetext}

\begin{figure}[htb]
\begin{center}
\scalebox{0.8}{\includegraphics{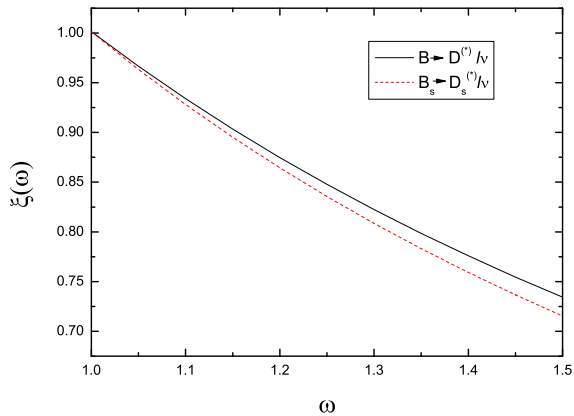}}
\end{center}
\caption{The dependence of Isgur-Wise function $\xi(\omega)$ on
$\omega$ with the typical values of $\Delta_{H}$=0.3 GeV and
$\Delta_{H}=0.5$ GeV corresponding to $B\to D^{(*)}l\nu$ and
$B_{s}\to D^{(*)0}_{s}l\nu$ decays respectively.\label{IS-1}}
\end{figure}

\begin{figure}[htb]
\begin{center}
\scalebox{0.8}{\includegraphics{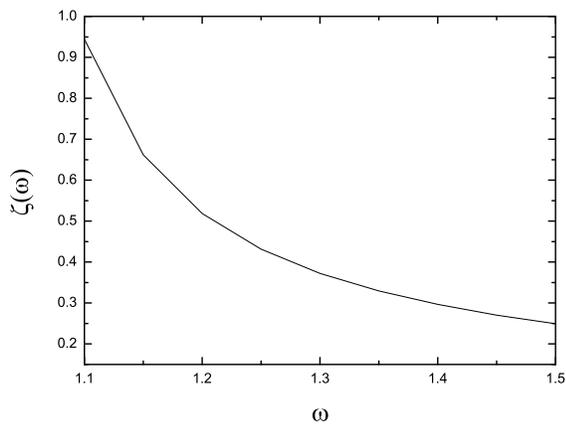}}
\end{center}
\caption{The dependence of $\zeta(\omega)$ on $\omega$ with the
typical values of $\Delta_{H}$=0.5 GeV and $\Delta_{S}=0.86$ GeV
corresponding to $B_s\to D_{sJ}(2317,2460)l\bar{\nu}$
decays.\label{zeta}}
\end{figure}

In Fig. \ref{IS-1}, we show the dependence of Isgur-Wise function
$\xi(\omega)$ on $\omega$ in $B\to D^{(*)}l\nu$ and $B_{s}\to
D^{(*)0}_{s}l\nu$ decays. The dependence of $\zeta(\omega)$ on
$\omega$ is shown in Fig. \ref{zeta}

For a comparison, we also put the values of the branching ratios of
$B_{s}\to D_{s}(1968)l\bar{\nu}$, $B_{s}\to
D_{s}^{*}(2112)l\bar{\nu}$, $B_{s}\to D_{sJ}^{*}(2317)l\bar{\nu}$
and $B_{s}\to D_{sJ}(2460)l\bar{\nu}$, which are calculated in Ref.
\cite{aliev1,aliev2}, in Table \ref{tab1}.
\begin{widetext}
\begin{center}
\begin{table}[h]
\begin{tabular}{c|ccccc}
  \hline
   processes & CQM & QSR in HQET ($m_{Q}\to \infty$)\cite{huang}&QSR   \\
  \hline\hline
  BR[$B_s\rightarrow D_{s}(1968)l\bar{\nu}$] &$(2.73\sim3.00)\%$ &-& - \\
  \hline
  BR[$B_s\rightarrow D^*_{s}(2112)l\bar{\nu}$] & $(7.49\sim7.66)\%$ & - & - \\
  \hline
  BR[$B_s\rightarrow D^*_{sJ}(2317)l\bar{\nu}$] & $(4.90\sim5.71)\times10^{-3}$ & $0.09$&
  $\sim 10^{-3} $ \cite{aliev1}\\
  \hline
  BR[$B_s\rightarrow D_{sJ}(2460)l\bar{\nu}$] &$(7.52\sim8.69)\times10^{-3}$
  &$0.08$& $4.9\times 10^{-3} $\cite{aliev2} \\
  \hline
\end{tabular}
\caption{In this table, we list our results of the semileptonic
decays of $B_{s}$ to $D_{s}(1968)$, $D_{s}^{*}(2112)$,
$D_{sJ}^{*}(2317)$ and $D_{sJ}(2460)$ and that obtained by other
approaches \cite{huang,aliev1,aliev2}. \label{tab1}}
\end{table}
\end{center}
\end{widetext}

\section{Discussion and conclusion}

By the comparison between the theoretical results and experimental
results listed in Table \ref{B-Dlnu}, we have reason to believe that
the CQM model can be applied well to study these semileptonic decays
related to this work. In this work, we study the semileptonic decays
of $B_s$ to $D_{s}(1968)$, $D_{s}^{*}(2112)$, $D_{sJ}^{*}(2317)$ and
$D_{sJ}(2460)$. In the HQET, $D_{s}(1968)$ and $D_{s}^{*}(2112)$ can
be well categorized as $H$ doublet ($0^{-},1^{-}$). To calculate the
semileptonic dedays relevant to $D_{sJ}^{*}(2317)$ and
$D_{sJ}(2460)$, we make an assumption that $D_{sJ}^{*}(2317)$ and
$D_{sJ}(2460)$ belong to $S$ doublet $(0^{+},1^{+})$.

At present the experiments only give $BR[B_{s}^{0}\to
D_{s}^{-}l^{+}\nu_{l}\mbox{anything}]=(7.9\pm2.4)\%$ \cite{PDG} and
the information of $B_{s}\to D_{s}(1968)l\bar{\nu}$ is still absent.
Thus the result of $B_{s}\to D_{s}(1968)l\bar{\nu}$ by the CQM model
does not contradict this experimental value. Meanwhile one predicts
the branching ratio of $B_{s}\to D_{s}^{*}(2112)l\bar{\nu}$ whose
order of magnitude is same as that of $B_{s}\to
D_{s}(1968)l\bar{\nu}$. $B_{s}\to D_{s}^{*}(2112)l\bar{\nu}$
semileptonic decay should be seen in future experiments. In Fig.
\ref{IS-1}, the dependence of Isgur-Wise function of $B_{s}\to
D^{(*)0}_{s}l\nu$ on $\omega$ is very similar to that of $B\to
D^{(*)}l\nu$. The difference between them comes from the breakings
of SU(3) symmetry and HQET symmetry. Thus we believe that the model
is applicable to our processes and expect to otain relatively
reliable results.

The branching ratios of $B_{s}\to D_{sJ}(2317,2460)l\bar{\nu}$ from
our calculations and that obtained by QCD sum rules
\cite{aliev1,aliev2} are of the same order of magnitude. However,
our predictions are far smaller than those given by Ref.
\cite{huang}. Anyway, at present both our calculations and the
analyses from other groups all indicate that the semileptonic decays
$B_{s}\to D_{sJ}^{*}(2317,2460)l\bar{\nu}$ own large branching
ratios. Therefore we urge our experimental colleagues to measure
those semileptonic channels in the CDF experiment and in the future
LHCb experiment. It will help us to further understand the nature of
those exotic $D_{sJ}$ mesons. And more future experiments will also
improve our understanding about the models applied to calculate the
semileptonic decays of $B_{s}$ to $D_{sJ}$ mesons.

\section*{Acknowledgment}

One of the authors, X.L., is indebted to Prof. Shi-Lin Zhu who
brought this topic to our attentions. We thank Prof. Xue-Qian Li for
useful discussions. X.L. was partially supported by the National
Natural Science Foundation of China under Grants 10375003 and
10421503, Key Grant Project of Chinese Ministry of Education (No.
305001) and the China Postdoctoral Science Foundation (No.
20060400376). S.M.Z. was supported by the Research Fund for the
Doctoral Programs of Hebei University.

\section*{Appendix}

The explicit expressions of $\mathcal{I}_{1,5}$ which are related to
our calculation are listed
%\begin{widetext}
\begin{eqnarray}
&&\mathcal{I}_{1}=\frac{iN_c}{16
\pi^4}\int^{reg}\frac{d^4k}{k^2-m_{s}^2}=\frac{N_c
m_{s}^2}{16\pi^2}\Gamma\left(-1,\frac{m_{s}^2}{\Lambda^2}
,\frac{m_{s}^2}{\mu^2}\right),\\
&&\mathcal{I}_{5}(\alpha_1,\alpha_2,\omega)\nonumber\\&&=\frac{iN_c}{16\pi^4}\int^{reg}\frac{\mathrm{d}^4k}{(k^2-m^2)(v\cdot
k+\alpha_1+i\epsilon)(v'\cdot k+\alpha_2+i\epsilon)}\nonumber\\
&&=\int_0^1\mathrm{d}x\frac{1}{1+2x^2(1-\omega)+2x(\omega-1)}\nonumber\\&&\times\bigg[
\frac{6}{16\pi^{3/2}}\int_{1/\Lambda^2}^{1/\mu^2}\!\!\mathrm{d}s\;\varrho\;
 e^{-s(m_{s}^2- \varrho^2)}s^{-1/2}(1+\mathrm{erf}(\varrho
 \sqrt{s}))\nonumber\\&&+\frac{6}{16\pi^2}\int_{1/\Lambda^2}^{1/\mu^2}\!\!\mathrm{d}s\;
  e^{-s(m_{s}^2- 2\varrho^2)}s^{-1}\bigg],
\end{eqnarray}
where the definitions of $erf(z)$ and $\varrho$ are
\begin{eqnarray}
&&\mathrm{erf}(z)=\frac{2}{\sqrt{\pi}}\int_{0}^z \mathrm{d}x
e^{-x^2},\\&&\varrho=\frac{\alpha_1(1-x)+\alpha_2x}{\sqrt{1+2(\omega-1)x+2(1-\omega)x^2}}.
\end{eqnarray}
%\end{widetext}

\end{document}